\documentclass[twocolumn]{aastex631}

\usepackage{natbib}
\usepackage{epsfig}
\usepackage{graphicx}
\usepackage{subfigure}
\usepackage{float}
\usepackage{amsmath}
\usepackage{color}
\usepackage{amssymb}
\usepackage{apjfonts}

\received{September 17, 2023}
\shortauthors{Liu \& Melia}

\graphicspath{{./}{figures/}}
\usepackage{amsmath}
\begin{document}

\title{Challenges to Inflation in the Post-Planck Era}

%% The \author command is the same as before except it now takes an optional
%% argument which is the 16 digit ORCID. The syntax is:
%% \author[xxxx-xxxx-xxxx-xxxx]{Author Name}

\correspondingauthor{Fulvio Melia}
\email{fmelia@email.arizona.edu}

\author{Jingwei Liu}
\affiliation{Department of Physics, The University of Arizona, AZ 85721, USA}

\author{Fulvio Melia \thanks{John Woodruff Simpson Fellow.}}
\affiliation{Department of Physics, The Applied Math Program and Department of Astronomy,
The University of Arizona, AZ 85721, USA}

\begin{abstract}
Space-based missions studying the cosmic microwave background (CMB) have progressively
refined the parameter space in conventional models of inflation shortly ($\sim 10^{-37}$ 
seconds) after the big bang. While most inflationary scenarios proposed thus far in the 
context of GR have since been ruled out, the basic idea of inflation may still be tenable, 
albeit with several unresolved conundrums, such as conflicting initial conditions and 
inconsistencies with the measured CMB power spectrum. In the new slow-roll inflationary 
picture, inflation arising in plateau-like potentials requires an initiation beyond the 
Planck time. This delay may be consistent with the cutoff, $k_{\rm min}$, measured 
recently in the primordial power spectrum. However, the actual value of $k_{\rm min}$ would 
imply an initiation time too far beyond the big bang for inflation to solve the horizon 
problem. In this paper, we also describe several other undesirable consequences of this 
delay, including an absence of well motivated initial conditions and a significant 
difficulty providing a viable mechanism for properly quantizing the primordial fluctuations. 
Nevertheless, many of these inconsistencies may still be avoided if one introduces 
non-conventional modifications to inflation, such as a brief departure from slow-roll 
dynamics, possibly due to a dramatic change in the inflationary potential, inflation 
driven by multiple fields, or a non-minimal coupling to gravity. In addition, 
some of these difficulties could be mitigated via the use of alternative cosmologies 
based, e.g., on loop quantum gravity, which replaces the initial big-bang singularity 
with finite conditions at a bounce-like beginning.
\end{abstract}


\section{Introduction}\label{introduction}
Inflationary cosmology was introduced several decades ago
\citep{Starobinskii:1979,Kazanas:1980,Guth:1981,Linde:1982}. The rapid expansion it
would have created in the early Universe might resolve several inconsistencies with
the standard model, including a solution to the cosmic microwave background (CMB)
temperature horizon problem, a dilution of the number density of magnetic monopoles to
undetectable levels today, and an explanation for why the Universe is spatially flat.
It may also have produced the primordial spectrum of fluctuations responsible for the
eventual formation of large-scale structure \citep{Mukhanov:1992}. But we still do not
know inflation's underlying field and potential, and have very little guidance
concerning how and when the inflationary phase may have started. Several other
difficulties still cause great concern regarding its ultimate viability, e.g., given
that inflation appears to conflict with several general relativistic constraints
\citep{Melia:2020} and, most seriously, violates at least one of the fundamental energy
conditions in Einstein's theory \citep{Melia:2023e}.

Our attention in this paper will be focused on the conventional picture of
inflation, i.e., the paradigm developed in the context of general relativity, based in
part on the Friedmann-Lema\^itre-Robertson-Walker metric (FLRW) defined in Equation~(\ref{eq:FLRW})
below. We shall examine how the various constraints now available to us, principally the
size of the homogeneous Universe today \citep{PlanckVI:2020}, the inferred very rapid 
expansion starting right at the big bang singularity, inflation's energy scale measured 
by {\it Planck}, and the apparent cutoff in the primordial fluctuation spectrum measured 
in the CMB \citep{MeliaLopez:2018,Melia:2021b,Sanchis-Lozano:2022}, cannot all be merged
together into a self-consistent picture. We must emphasize at the outset, however, that
alternative early universe scenarios have been proposed in recent years, in part to
circumvent at least some of the problems we are highlighting in this paper. Though
a period of inflation is still required in these models, the conditions leading up
to the initiation of the exponential expansion can be drastically different from those
in the conventional picture. 

For example, a cosmology based on loop quantum gravity 
\citep{Bojowald:2008,Ashtekar:2003,Ashtekar:2011}---a modified version of general 
relativity that incorporates the Heisenberg uncertainty principle from quantum mechanics 
on scales approaching the Planck domain (see Eqs.~\ref{eq:lPl} and 
\ref{eq:rhoPl})---replaces the initial singularity with a bounce 
\citep{Ashtekar:2006,Martin-Benito:2009} and a small Hubble constant with a 
corresponding gravitational radius \citep{Melia:2018b} much larger than all of the other 
length scales. As we shall discuss more extensively in \S~\ref{discussion}, these changes 
can greatly mitigate some of the more serious defects in the standard inflationary paradigm
\citep{Taveras:2008,Bhardwaj:2019,Ashtekar:2020,Navascues:2021}. Our analysis will 
not be directed towards such models.

But insofar as the conventional inflationary picture is concerned, prior to the 
{\it Planck}-2013 data release \citep{Planck:2014}, the basic paradigm was the 
so-called classic ``chaotic'' inflation \citep{Linde:1983}, in which a
minimally-coupled scalar (inflaton) field, $\phi$, emerged at the Planck time,
$t_{\rm Pl}$, with an energy density set by the Planck scale (see Eq.~\ref{eq:rhoPl}
below).  As the precision of the CMB measurements continued to improve, however, particularly
with the most recent {\it Planck} analysis \citep{PlanckVI:2020}, this simple picture
has not continued to fare well.

The principal reason for this is that, after the {\it Planck}-2013 release, the simplest 
inflationary scenarios have been disfavoured in comparison with a particular class
of models with a plateau-like potential, $V(\phi)$ \citep{Planck:2014,Ijjas:2013,Ijjas:2014}.
But in such cases, $V(\phi)$ has a minimum at the end of inflation, and must
have changed slowly with $\phi$ at the beginning, which precludes any possibility of
the Universe's energy density matching the Planck value at $t_{\rm Pl}$, unless the
de~Sitter expansion was initiated at $t_{\rm init}\gg t_{\rm Pl}$.
This undesirable behavior---which motivates the `new' chaotic inflationary picture---is
unavoidable when the various observables are fit to the data, especially the tensor to scalar
ratio, ${\mathcal{R}}$, and the spectral index, $n_{\rm s}$, of the scalar fluctuation spectrum
\begin{equation}
P_{\rm s}(k)\delta ( \textbf {k}+\textbf{k}')(2\pi)^3 = \langle
\Theta_{\textbf{k}}\Theta_{\textbf{k}'}\rangle\label{eq:Pk}
\end{equation}
\vfill\eject\noindent
or, more accurately, the dimensionless power spectrum
\begin{equation}
\Delta_{\rm s}^2\equiv {k^3\over 2\pi^2}P_{\rm s}(k)\;,\label{eq:Deltak}
\end{equation}
where
\begin{equation}
n_{\rm s}-1={d \ {\rm ln}\ \Delta_{\rm s}^2\over d \ {\rm ln} \ k}\;.\label{eq:ns}
\end{equation}
Throughout this paper, $k$ is the comoving wavenumber of each mode, such that
\begin{equation}
k={2\pi a(t)\over \lambda_k}\;,\label{eq:k}
\end{equation}
in terms of its proper wavelength, $\lambda_k$, and the expansion factor, $a(t)$,
in the FLRW metric.

As we shall examine in this paper, however, the delayed initiation of the de Sitter phase
creates several major problems with the conventional picture of inflation
in the context of GR, the most significant of which is that its initial conditions
\citep{Guth:1981,Linde:1982} are then left unspecified \citep{Ijjas:2013,Ijjas:2014}.
That is, why was the inflaton field homogeneous over distances spanning causally-disconnected
regions?  Slow-roll inflation needs $\phi$ to have been homogeneous when the slow-roll phase
started at $t_{\rm init}$. This is required by the condition that the kinetic and
spatial-derivative terms of $\phi$ be negligible compared to $V(\phi)$.

Furthermore, the primordial fluctuations in $\phi$ would themselves have been seeded
at $t_{\rm init}$. But how then were they quantized? Such a delayed starting
time would not have allowed all the modes to be seeded well below the gravitational
horizon, $R_{\rm h}\equiv c/H$ (i.e., the Hubble radius), to satisfy the Bunch-Davies
vacuum conditions. These hurdles are in addition to the more obvious challenge of addressing
the horizon problem when inflation is delayed too far beyond the big bang. For these
reasons, the new chaotic inflation is far from established as a viable paradigm of the
early Universe (see also \citealt{Melia:2022a} for a more detailed description 
of the current problems faced by the standard model).

In this paper, we shall first affirm the growing consensus that the most recent
data unavoidably require $t_{\rm init}\gg t_{\rm Pl}$ in the context of slow-roll,
plateau-potential models (\S~\ref{background}). Then we examine how and why an
inflationary expansion beginning at $t_{\rm init}\gg t_{\rm Pl}$ necessitates a
sharp cutoff $k_{\rm min}$ in the primordial power spectrum (\S~\ref{kmin}),
possibly explaining the large-angle anomaly seen in the CMB anisotropies. But
we then also explain why the delay creates the negative features described above
(\S~\ref{Other}). In our discussion, we shall include a brief survey of how some 
of these difficulties may be circumvented in alternative theories, such as loop 
quantum cosmology, and we end with our closing thoughts in \S~\ref{conclusion}.
\vfill\newpage

\section{Background}\label{background}
We adopt the FLRW metric for the cosmic spacetime, written
in terms of the comoving coordinates $(ct,r,\theta,\phi)$, and the spatial curvature
constant, $K$ (to properly distinguish it from the more commonly used
symbol $k$ denoting the mode wavenumber):
\begin{equation}
ds^2=dt^2-a(t)^2\left[{dr^2\over 1-Kr^2}+r^2\left(d\theta^2
+\sin^2\theta\,d\phi^2\right)\right].\label{eq:FLRW}
\end{equation}
The observations suggest that the Universe is probably spatially flat, so we shall
simply set $K=0$ throughout this paper (see also \citealt{Melia:2022a}).
The universal expansion factor, $a(t)$, evolves according to the imposed
equation-of-state in the stress-energy tensor and, given that we
adopt spatial flatness throughout this paper, we normalize it to $1$ at $t=t_0$,
where $t_0$ is the current age of the Universe.

Our classical description of the Universe breaks down prior to the 
Planck time, $t_{\rm Pl}\equiv l_{\rm Pl}/c$, corresponding to the Planck spatial 
scale 
\begin{equation}
l_{\rm Pl}=\sqrt{{2Gh\over c^3}}\;.\label{eq:lPl}
\end{equation}
Numerically, we have $l_{\rm Pl}\approx 5.7\times 10^{-33}$ cm and
$t_{\rm Pl}\approx 1.9\times 10^{-43}$ s. The Planck energy density is therefore
simply given as
\begin{equation}
\rho_{\rm Pl}\equiv {3m_{\rm Pl}c^2\over4\pi l_{\rm Pl}^3}\;.\label{eq:rhoPl}
\end{equation}

As noted earlier, a significant challenge now faced by the `new' 
chaotic inflationary paradigm in the context of GR is that the latest {\it Planck} 
measurements preclude the actual density in the early Universe from matching 
Equation~(\ref{eq:rhoPl}) with a slow-roll inflaton potential. To make the discussion 
quantitative, we shall focus on the Higgs-like potential and demonstrate how it fails 
to satisfy the observational constraints.

This potential has the form:
\begin{equation}
V=V_0\left[1-\left({\phi \over \mu}\right)^2\right]^2\;,\label{eq:Higgslike}
\end{equation}
from which we can determine the so-called slow-roll parameters,
\begin{equation}
\epsilon_V\equiv {M_{\rm Pl}^2 \over 2}\left({V_{,\phi} \over V}\right)^2
={8M_{\rm Pl}^2\phi^2\over\mu^4[1-({\phi / \mu})^2]^2}\label{eq:HiggsEpsilon}
\end{equation}
and
\begin{equation}
\eta_V\equiv M_{\rm Pl}^2{V_{,\phi\phi} \over V}
=-{4M_{\rm Pl}^2(\mu^2-3\phi^2)\over(\mu^2-\phi^2)^2}\;,\label{eq:HiggsEta}
\end{equation}
where $V_{,\phi}$ denotes the first derivative of $V$ with respect to $\phi$, and
$V_{,\phi\phi}$ is its second derivative. In these expressions, $M_{\rm Pl}$ is
the reduced Planck mass, defined as $M_{\rm Pl} = {m_{\rm Pl}}/{\sqrt{8\pi}}$.
Under the slow-roll approximation, we obtain:
\begin{equation}
n_{\rm s}-1 = 2\eta_V-6\epsilon_V\;
\end{equation}
and
\begin{equation}
{\mathcal{R}} = 16\epsilon_V\;.
\end{equation}
Adopting the {\it Planck} measurement, $n_{\rm s}=0.966$, and the upper 
limit of the tensor to scalar ratio, ${\mathcal{R}}=0.036$ \citep{Ade:2021}, we find 
that $\mu \approx 18.7\;M_{\rm Pl}$ and $\phi_{0.002} \approx 5.4\;M_{\rm Pl}$,
denoting the value of $\phi$ measured at $k=0.002$ Mpc$^{-1}$. The fact that 
$({\phi_{0.002}/\mu})^2\ll1$ confirms our inference that $V$ would have been very 
nearly constant at the early stage of inflation. As long as $V(\phi)$ dominated the 
energy density of the Universe at that point, we conclude that $H$ must also have been
approximately constant during the early stages of inflation.

The dynamical evolution of $\phi$, however, must allow its energy density $\rho(\phi)$
($\propto H^2$) to self-consistently match $\rho(t_{\rm Pl})$ at $t_{\rm Pl}$. With
slow-roll \citep{Linde:1983},
\begin{equation}
\Delta_{\rm s}^2 = {H^2\over {8 \pi^2 M_{\rm Pl}^2 \epsilon}}={2H^2\over
{\pi^2 M_{\rm Pl}^2 {\mathcal{R}}}}\;.\label{eq:Deltak2}
\end{equation}
But from the definition of the amplitude, $A_{\rm s}$, of the primordial 
power spectrum, we also have
\begin{equation}
\Delta_{\rm s}^2 = A_{\rm s}\left({k \over k_*}\right)^{n_{\rm s}-1}\;,\label{eq:As}
\end{equation}
where $k_*$ is a pivot scale, usually taken to be $0.05$ Mpc$^{-1}$. 
Thus, inserting the Planck measured value of the amplitude, $A_{\rm s} = 2.1 \times 10^{-9}$,
we conclude that $H_{\rm init} \approx H_{0.002}=3.1 \times 10^{37}$ s$^{-1}$.

From the Friedmann equation (with $K=0$), we estimate a density
\begin{equation}
\rho(a_{\rm init})={3c^2\over 8\pi G}H(a_{\rm init})^2\;.\label{eq:Friedmann}
\end{equation}
Numerically, it appears that $\rho(a_{\rm init}) \approx 1.7 \times 10^{84}$ kg m$^{-3}$.
But from Equation~(\ref{eq:rhoPl}), we have instead $\rho(a_{\rm Pl}) \approx 5.2
\times 10^{96}$ kg m$^{-3}$, representing a significant difference of $\sim 10^{12}$
in the density between these two times.  The dynamical evolution of $\phi$, which
largely depends on $V(\phi)$, must allow sufficient time to pass ($a_{\rm init} \gg
a_{\rm Pl}$) for $\rho(\phi)$ to drop well below its Planck value.

\section{Truncation of the primordial power spectrum }\label{kmin}
On its own, the fact that $t_{\rm init}\gg t_{\rm Pl}$ already creates a
significant challenge in ensuring self-consistency between $V(\phi)$ and
the physical conditions in the early Universe. But a recent analysis
of the CMB anisotropies measured by {\it Planck} provides a more severe
constraint on the possible value of $a_{\rm init}$ itself. This work
demonstrates in a model-independent way how the angular correlation
function of the temperature distribution unavoidably points to a distinct
cutoff, $k_{\rm min}$, in the primordial power spectrum
\citep{MeliaLopez:2018,Melia:2021b,Sanchis-Lozano:2022}.

To be clear, this inference that the measured angular correlation
function of the CMB is best explained with a clean cutoff, $k_{\rm min}$, is
specifically associated with the conventional inflationary paradigm. Several
other workers have probed the possibility that the missing angular correlation
on large scales (and the missing power at small multipole moments) may instead
be due to dynamical effects. Again using loop quantum cosmology as an evident
example, only modes with the largest observable wavelengths in such models would
have felt the quantum curvature effects near the bounce \citep{Agullo:2015,Gomar:2017,Ashtekar:2017}, 
undergoing a pre-inflationary phase of kinetic dominance that eventually evolved 
into the standard slow-roll profile. This sequence of steps can effectively 
suppress their power relative to the smaller $k$ modes, thereby alleviating 
the CMB anomalies \citep{MartindeBlas:2016,Ashtekar:2020}. 

At first sight, this evidence of a cutoff, $k_{\rm min}$, is actually quite exciting 
in the context of a delayed initiation of inflation in standard GR. In the analysis of 
the CMB anisotropies \citep{Hinshaw:1996,Bennett:2003,PlanckVI:2020}, the tacit assumption 
is usually made that quantum fluctuations in the inflaton field started exiting
the Hubble horizon at $t < t_{\rm Pl}$, effectively implying no lower cutoff in
the primordial power spectrum $P_{\rm s}(k)$ since $k_{\rm min}$ would then have been much
smaller than any modes observable in the CMB today (see \citealt{Melia:2020} and references
cited therein). But the more recent analysis of the {\it Planck} data, focusing on the value
of $k_{\rm min}$ itself, has revealed that this simple picture is not consistent with the
absence of large-angular correlations in the CMB temperature anisotropies
\citep{LiuMelia:2020}.

\begin{figure*}[ht!]
\centering
\includegraphics[width=120mm]{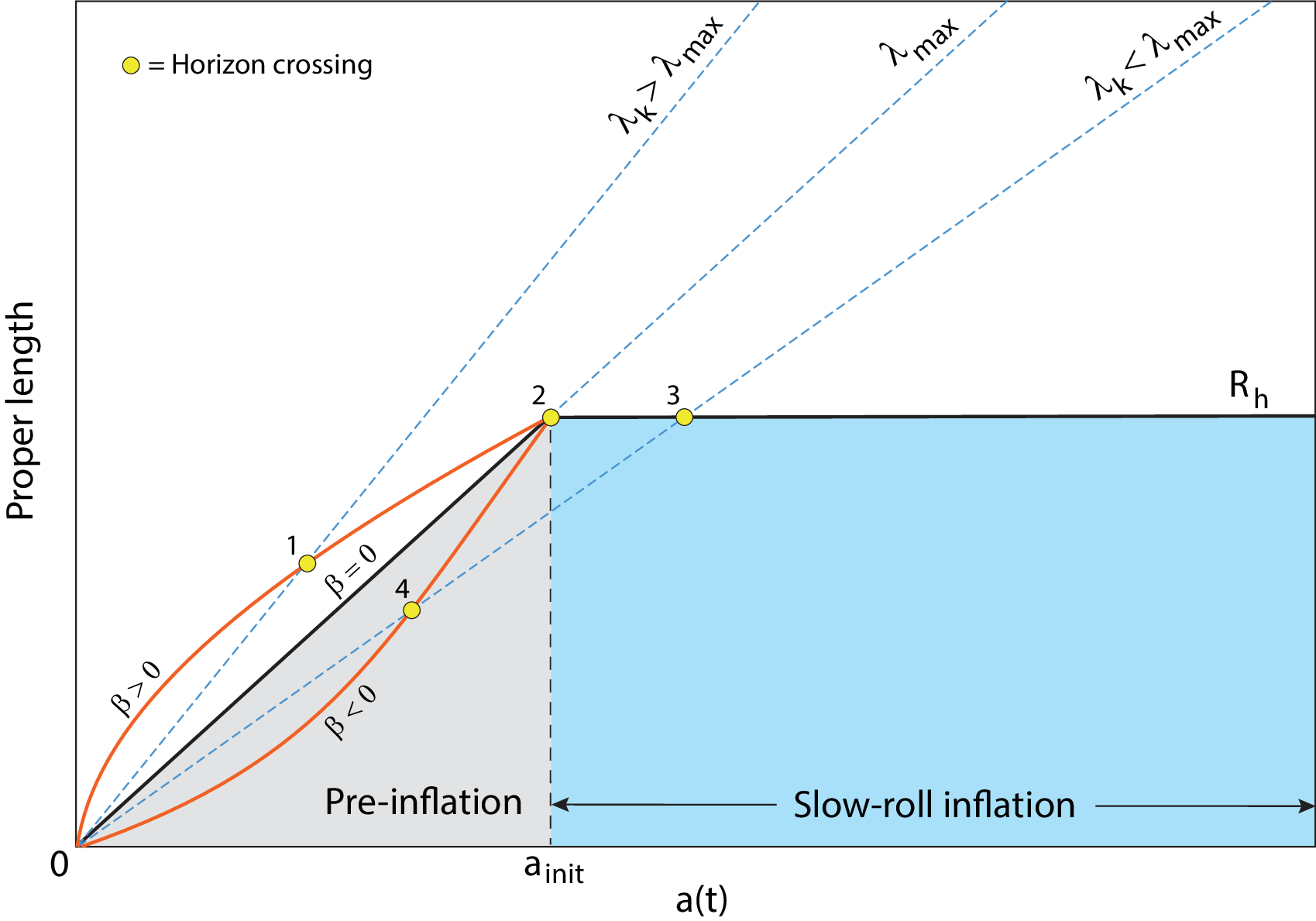}
\caption{Proper length as a function of the universal expansion factor $a(t)$, for
the Hubble radius $R_{\rm h}\equiv c/H$ (solid, black when $\beta=0$; red curves
when $\beta\not=0$), mode wavelength ($\lambda_k$) and maximum mode wavelength
($\lambda_{\rm max}$) (straight dashed lines). Slow-roll inflation (blue shaded region)
initiates at $a_{\rm init}\equiv a(t_{\rm init})$, with $t_{\rm init}\gg t_{\rm Pl}$.
The linear pre-inflationary phase is indicated by the grey shaded region. The yellow
circles illustrate several horizon crossings by various modes.}
\label{fig:Fig1}
\end{figure*}

Three independent missions \citep{Hinshaw:1996,Bennett:2003,PlanckVI:2020} have now confirmed
the lack of angular correlation in the CMB temperature profile at angles $\gtrsim 60^\circ$.
The most likely explanation for this anomaly appears to be the presence of a hard cutoff,
\begin{equation}
k_{\rm min}=(3.14 \pm 0.36)\times 10^{-4}\;{\rm Mpc ^{-1}}\;,\label{eq:kmin}
\end{equation}
in the primordial power spectrum. As we shall see shortly (\S~\ref{Other}), this cutoff
implies a specific time at which modes started exiting the horizon, and this is clearly
well beyond the Planck time.

The interesting aspect of this finding is that---contrary to the prediction of
classic chaotic inflation---a cutoff is unavoidable when $a_{\rm init}\gg a_{\rm Pl}$.
The issue is then whether the `measured' value of $k_{\rm min}$ consistently yields the
same initiation time $t_{\rm init}$ as that implied by the argument we have summarized
in \S~\ref{background}.

Here, we shall demonstrate quantitatively why a delayed initiation for inflation must
produce a truncated primordial power spectrum in the context of slow-roll models
with a plateau-like potential, since these are the most commonly used scenarios.
The almost scale free nature of the primordial power-spectrum implies that, during
inflation, the Hubble radius $R_{\rm h}={c/ H}$ (and thus the Hubble parameter
$H$) changes very slowly.

The physical reason for this is not difficult to understand. Quantum fluctuations
in $\phi$ oscillate and decay as long as their wavelength $\lambda_k\equiv
2\pi a(t)/k$ is smaller than $2\pi R_{\rm h}$. Once $\lambda_k>2\pi R_{\rm h}$,
however, they stop oscillating and their amplitude tends toward a constant value
\citep{Bardeen:1983}.  Only their wavelength continues to change (in proportion to
$a$) as the Universe expands.  Horizon crossing corresponds to that moment at which
$\lambda_k=2\pi R_{\rm h}$, and is thus an essential ingredient in the generation of 
observable density fluctuations in inflationary cosmology. A perfectly scale-free 
primordial spectrum would correspond to modes crossing $R_{\rm h}$ with a constant 
$H$ value. The fact that $n_{\rm s}$ is not exactly one (but very close to it) implies 
that $H$ is not exactly constant, but can vary only slightly during the time when the 
spectrum is created.

In this picture, the larger a mode is, the smaller is its wavenumber $k$, and
therefore the earlier is its exit across the horizon. For simplicity, we consider
the simple pre-inflationary scaling
\begin{equation}
H(a)=Ba^{-1+\beta}\qquad (t\le t_{\rm init})\;,\label{eq:Hpreinf}
\end{equation}
which leads to
\begin{equation}
H(t)={1\over{(1-\beta)t}}\;,\label{eq:Htpre}
\end{equation}
and
\begin{equation}
\rho(t_{\rm Pl})={3\over 8\pi G (1-\beta)^2 t_{\rm Pl}^2}\;.\label{eq:rhotPl}
\end{equation}
A comparison of Equation~(\ref{eq:rhotPl}) with (\ref{eq:rhoPl}) shows that
$\rho(t_{\rm Pl})$ matches the Planck energy density exactly when $\beta=0$
and is very close to it as long as $|\beta|\ll 1$. There is thus an additional
motivation for considering this type of scaling in the pre-inflationary Universe,
since it provides a `natural' way for the measured energy of inflation to
merge smoothly toward the Planck scale. The mismatch in densities would
be difficult to address otherwise. In any case, Equation~(\ref{eq:Hpreinf})
is quite general because it could also represent a radiation-dominated universe
(RD), with $H \propto a^{-2}$, a kinetic-dominated universe (KD) 
\citep{Contaldi:2003}, with $H \propto a^{-3}$ \citep{LiuMelia:2020}, 
and even a string-dominated universe (SD), with $H \propto a^{-1}$ \citep{Spergel:1997}. 
Interestingly, the latter option would yield the simplest, most compelling solution for
the energy density mismatch between $t_{\rm Pl}$ and $t_{\rm init}$.

The growth of different modes and the Hubble radius then follow the trajectories
illustrated in Figure~\ref{fig:Fig1}. The yellow dots represent typical Horizon
crossing times. As we can see , for all the cases, horizon crossing can happen
during inflation (point 3), but for $\beta \le 0$, it cannot happen before the
inflation. Indeed, there actually exists a largest possible mode ($\lambda_{\rm max}$)
whose crossing point corresponds to the beginning of inflation (at $a_{\rm init}$).
Horizon crossing prior to inflation can only occur for $\beta > 0$. Even here,
however, we would still not see a primordial power spectrum without $k_{\rm min}$
due to the requirements imposed by the quantization of the seed fluctuations (see
below). In this regard, notice that all of the three examples we pointed to, RD, KD
and SD, have $\beta \le 0$. It is thus clear that all the simple and well motivated
models one can consider in the delayed initiation picture lead to the existence of
a hard cutoff $k_{\rm min}$ in the primordial power spectrum.

\section{Challenges due to the delayed initiation time}\label{Other}
The cutoff $k_{\rm min}$ corresponds to a specific time $t_{\rm init}$ (and the
corresponding $a_{\rm init}$) at which the larget mode crossed the Hubble horizon.
To estimate it, we consider the null condition in Equation~(\ref{eq:FLRW}), for a
photon propagating radially between times $t_1$ and $t_2$:
\begin{equation}
r=c\int_{t_1}^{t_2}{dt\over a}\;.\label{eq:r}
\end{equation}
Using the Hubble parameter,
\begin{equation}
H={\dot{a}\over a}\;,\label{eq:H}
\end{equation}
we may write
\begin{equation}
r_{\rm dec}=c\int_{a_{\rm cmb}}^{a_0}{da\over {a^2H}}\;,
\end{equation}
for the comoving distance traveled by a photon from decoupling to us, and
\begin{equation}
r_{\rm pre-cmb}=c\int_{a_{\rm Pl}}^{a_{\rm cmb}}{da\over {a^2H}}\;,
\end{equation}
for the corresponding quantity from the Plank time to decoupling.

We can estimate $r_{\rm dec}$ numerically using the Friedmann equation,
\begin{equation}
H^2=H_0^2 \left({\Omega_{\rm m}\over {a^3}} + {\Omega_{\rm r}\over {a^4}}+
{\Omega_\Lambda}\right),\label{eq:Friedmann2}
\end{equation}
where $\Omega_{\rm m}$, $\Omega_{\rm r}$ and $\Omega_\Lambda$ are the matter,
radiation and cosmological-constant energy densities, respectively, scaled to the
critical density today, $\rho_{\rm c}\equiv 3c^2H_0^2/8\pi G$. The result
is $r_{\rm dec} \approx 13,804$ Mpc \citep{LiuMelia:2020}, assuming the
{\it Planck} optimized parameters in the standard model \citep{PlanckVI:2020}.

The comoving distance $r_{\rm pre-cmb}$ may be decomposed into the contribution before
inflation ($r_{\rm pre-inf}$), during inflation ($r_{\rm inf}$) and after inflation (all
the way up to decoupling). It is not difficult to see, however, that the dominant
contribution is made prior to the end of the inflation. Furthermore, if we restrict
our attention to slow-roll potentials, we may approximate $r_{\rm inf}$ by setting
$a(t)$ equal to a simple exponential, the result of which is
\begin{equation}
r_{\rm inf}={c\over {a_{\rm init} H_{\rm init}}}\label{eq:rinf}
\end{equation}
where, as usual, the subscript `init' refers to the time at which inflation began.

To solve the horizon problem, we need
\begin{equation}
r_{\rm preCMB} \approx r_{\rm inf}>2r_{\rm dec}\approx 27,608{\rm Mpc}\;.\label{eq:2rdec}
\end{equation}
Thus, Equations~(\ref{eq:rinf}) and (\ref{eq:2rdec}), with the measured upper limit
on $H_{\rm init}$, give us $a_{\rm init}\lesssim 7.5\times 10^{-57}$.

But $a_{\rm init}$ must also correspond to the initial crossing time of the modes
\citep{LiuMelia:2020}, and the value of $t_{\rm init}$ corresponding to $k_{\rm min}$
is not necessarily constrained in the same way as the argument based on $H_{\rm init}$.
Looking again at Equation~(\ref{eq:rinf}), we now see that when $k_{\rm min}\not=0$,
the starting point of inflation can be found using the crossing condition of the largest mode
(see Eq.~\ref{eq:k}):
\begin{equation}
a_{\rm init} H_{\rm init}=c k_{\rm min}\;.\label{eq:ainit}
\end{equation}
Let us now compare the value $a_{\rm init}\lesssim 7.5\times 10^{-57}$ required to solve
the horizon problem with that obtained using $H_{\rm init}=H_{0.002}$ and $k_{\rm min}$.
Substituting these quantities into Equation~(\ref{eq:ainit}), we estimate that
$a_{\rm init}=9.7\times 10^{-56}$, too large by over an order of magnitude. We can
see this more concretely by estimating the corresponding comoving distances.
From Equations~(\ref{eq:rinf}), (\ref{eq:kmin}) and (\ref{eq:ainit}), we find
that $r_{\rm inf}\approx 3,181$ Mpc, which is much smaller than $2 r_{\rm dec}\approx
27,608$ Mpc. Thus, in order to solve the horizon problem while still maintaining
consistency with the presence of the measured $k_{\rm min}$, one must rely on the
pre-inflationary phase (i.e., $r_{\rm pre-inf}$) to make up the difference.

Any self-consistent picture for establishing the initial conditions for inflation must
therefore also be able to help inflation overcome the horizon problem. Of course, this
is not the only challenge, because the existence of $k_{\rm min}$, which implies that
$t_{\rm init}\gg t_{\rm Pl}$, creates several other potential issues. At the very
least, the pre-inflationary phase must account for the homogeneity of the
inflaton field at the beginning of the slow-roll phase. And the framework for allowing
this to happen must also provide a mechanism for properly quantizing the seed fluctuations
in $\phi$.

It doesn't take long to realize that a delayed initiation time for inflation merely
pushes several problems that inflation was supposed to solve farther back into the
pre-inflationary phase. The delayed initation time leaves the initial condition of
inflation unspecified. Slow-roll models require the Universe to be dominated by a
homogeneous inflaton field at the beginning of inflation, because they require
the kinetic and spatial-derivative terms of $\phi$ to be negligible compared to
$V(\phi)$. In other words, the homogeneous region at $t_{\rm init}$ must be large
enough to expand into the CMB sphere we see today.

As a result of this, the comoving distance a photon travelled before inflation, $r_{\rm preinf}$,
should be longer than $2r_{\rm dec}\approx 27,608$ Mpc. We can see if this is feasible
using the simple pre-inflationary scaling we introduced in \S~\ref{kmin}. Combining
Equations~(\ref{eq:Hpreinf}), (\ref{eq:r}) and (\ref{eq:H}), we have
\begin{equation}
r_{\rm preinf} = {c\over (-\beta)B}\left({1\over a_{\rm init}^\beta}-{1\over
a_{\rm Pl}^\beta}\right)\;.
\end{equation}
For $\beta<0$ (which, as we have seen, is the better motivated scenario), the fact that
$a_{\rm init}\gg a_{\rm Pl}$ implies that
\begin{equation}
r_{\rm preinf} = {c\over (-\beta)B a_{\rm init}^\beta}={c\over (-\beta)
H_{\rm init}a_{\rm init}}\;.
\end{equation}
And combining this with Equation~(\ref{eq:ainit}), with the measured value of
$k_{\rm min}$, we find that $|\beta|$ needs to be smaller than $0.13$ to ensure
that $r_{\rm preinf}>2r_{\rm dec}$. This rules out many models, including the
kinetic-dominated and radiation-dominated expansions. Furthermore, if inflation
is delayed to a time when the inflaton field was already homogeneous, then
the horizon problem would already have been solved during the pre-inflationary
phase. This would raise the bigger question of why we would then even
need inflation.

An equally big challenge would be to understand how the seed fluctuations could
have been quantized in this scenario. As we shall discuss in
\S~\ref{discussion}, the principal difficulty with the normalization of 
these modes is due to the time dependence of their frequency, which arises
from the changing spacetime curvature as they evolve. Selecting an otherwise
seemingly random time to canonically quantize them is therefore poorly
motivated, except in some alternative cosmologies that predict a bounce
instead of an initial singularity. One may thus find reasons to justify a
normalization at that initial time.

In the standard inflationary paradigm based on GR, canonical quantization can only 
occur in Minkowski space, where the frequency of the modes is time-independent. Of
course, the cosmic spacetime never satisfies this condition if $\ddot{a}\not=0$, but
with classical chaotic inflation, one could bypass this problem by assuming that the
fluctuations first emerged far enough into the conformal past, well below the Planck
scale (i.e., the `Schwarzschild' radius at the Planck time) where the spacetime curvature
could be neglected in estimating the mode amplitude. Known as the `Bunch-Davies'
vacuum \citep{Bunch:1978}, this remote patch of space has allowed classical chaotic
inflation to provide a means of identifying the quantum modes in $\phi$.

But this could be done in the classic ``chaotic'' picture because one could always
find an early enough time, $t\ll t_{\rm Pl}$, when $\lambda_k$ was much smaller
than the Hubble radius, $R_{\rm h}$, of the Universe (corresponding to its
gravitational horizon; \citealt{Melia:2018b}.) With the new chaotic inflationary picture,
however, the single field slow-roll phase beginning at $t_{\rm init}\gg t_{\rm Pl}$
no longer allows for this possibility. The larger modes, in particular, would never
have transitioned through a Bunch-Davies vacuum, preventing us from establishing
their canonical quantization condition.

\section{Discussion}\label{discussion}
All of our analysis thus far has been based on classic chaotic inflation. Not surprisingly,
however, at least some of these inconsistencies may be mitigated with the introduction
of unconventional modifications to the basic inflationary picture, with the effect of producing
additional degrees of freedom. For example, a helpful change would be the assumption
of a brief departure from slow-roll dynamics near the beginning of inflation. This could
be due to various factors, including an ad hoc dramatic shift in the inflationary potential
\citep{Hunt:2007,Qureshi:2017,Ragavendra:2022}, an early phase of inflation driven by multiple 
fields \citep{Braglia:2020}, or a non-minimal coupling to gravity \citep{Tiwari:2023}. 

In reality, these types of modification do not completely solve the problems 
faced by inflation in the conventional context of GR either. For instance, in the inflationary 
picture, the strong energy condition (SEC) from general relativity is violated during the 
slow-roll phase \citep{Melia:2023e}. Adding a brief non-slow-roll period at the beginning 
cannot eliminate this difficulty.

\begin{figure}
\centering
\includegraphics[width=85mm]{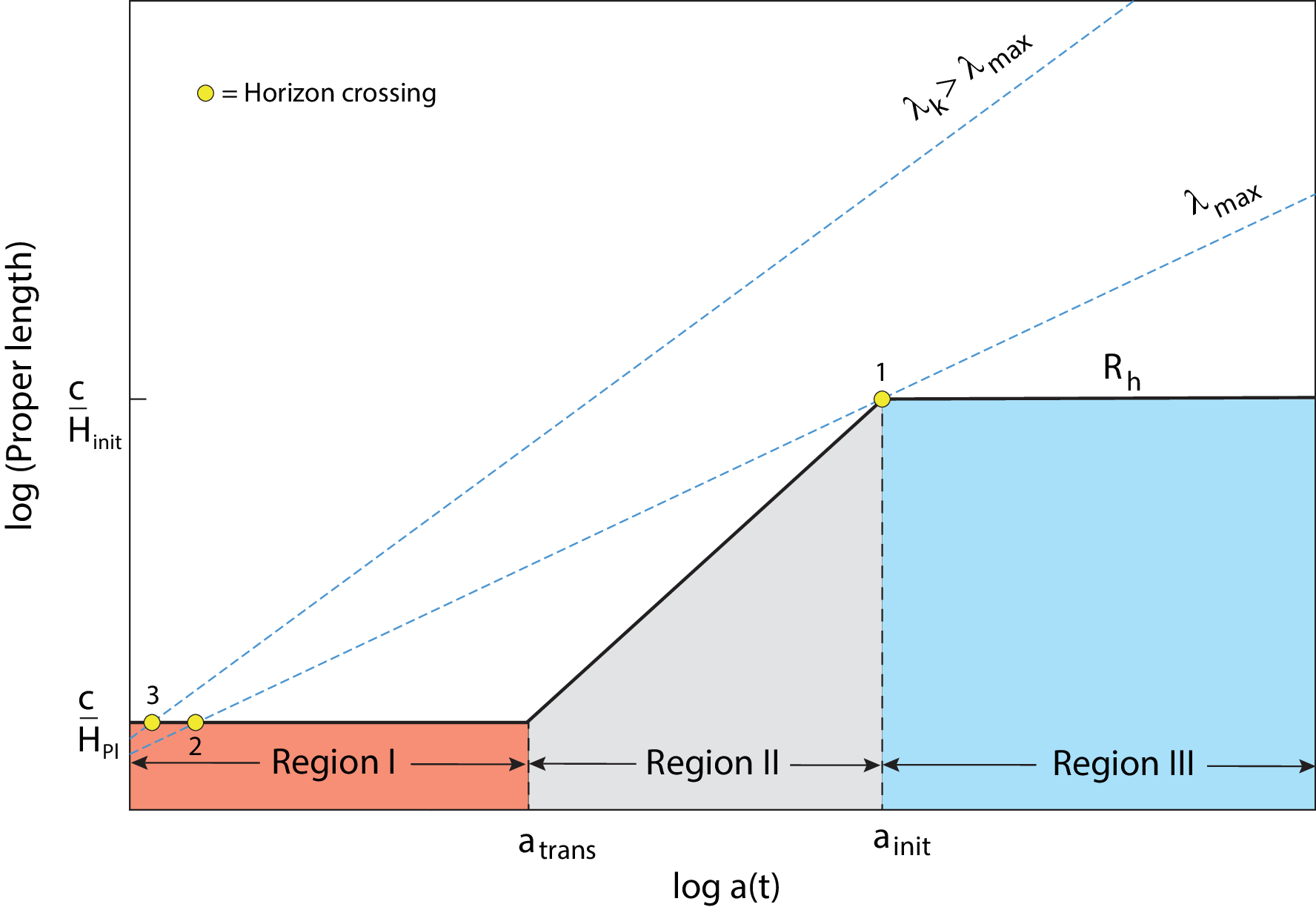}
\caption{Proper length as a function of $a(t)$: Hubble radius $R_{\rm h}\equiv c/H$ (solid, 
black); wavelength of the mode ($\lambda_{\rm max}$) corresponding to $k_{\rm min}$ (straight, 
dashed), crossing the horizon at points 1 and 2; wavelength of another mode leaving the horizon 
earlier than $\lambda_{\rm max}$, at point 3. The early de-Sitter expansion (red Region I) 
ends at $a_{\rm trans}$, where the kinetic dominated (KD) phase (grey Region II) begins. 
The KD expansion ends at $a_{\rm init}$, where slow-roll inflation (blue Region III) begins.}
\label{fig:Fig2}
\end{figure}

But to gauge how effective these changes may be in resolving the various issues we
have raised in this paper, it helps to consider a simplified scenario that embraces many
or all of the features required to make inflation work consistently with the data. To summarize,
the minimal set of requirements include the following: (i) The inflaton field must be present
sufficiently early (i.e., well below the Planck time) for its fluctuations to evolve through
the Bunch-Davies vacuum for canonical quantization; (ii) the inflation energy, characterized
by $H(a_{\rm init})$ at the initiation of the `new' chaotic slow-roll expansion, must be
matched to the Planck energy, characterized by $H(a_{\rm Pl})$, via the appropriate evolution
of $V(\phi)$ prior to $a_{\rm init}$ and the expansion history it generates; (iii) the entire
inflationary phase, beginning with $\phi$ in the Bunch-Davies vacuum, must produce a
sufficient number of e-folds to solve the horizon problem; and (iv) the fluctuation spectrum
must display a clear cutoff at $k_{\rm min}$, as discussed above. 

The schematic diagram in Figure~\ref{fig:Fig2} shows the evolution of proper distance as a 
function of the expansion factor $a$ consistent with all four of these minimal conditions. 
To produce this kind of expansion history, the inflaton potential, $V(\phi)$, must be
highly dynamic at $a\lesssim a_{\rm init}$, consistent with the various proposals
referenced earlier in this section. For example, the energy density must drop sharply
between $a_{\rm Pl}$ and $a_{\rm init}$, corresponding to roughly 12 orders of magnitude
in $H$, producing the steep gradient in $R_{\rm h}$, from $c/H_{\rm pl}$ to $c/H_{\rm init}$,
during this time. In other words, the slow-roll phase cannot start until $a_{\rm init}$,
which itself must correspond to the time at which the largest mode wavelength crosses
the horizon (at point 1 in this figure).

As we have seen, however, the observed value of $k_{\rm min}$ precludes the slow-roll
expansion from solving the horizon problem on its own. And a simple KD phase (Region II)
prior to that phase is not sufficient either. Thus, $V(\phi)$ must produce an earlier 
de Sitter (or quasi-de Sitter) expansion (Region 1) prior to $a_{\rm trans}$. This will
solve the horizon problem, effectively splitting the impact of inflation in producing
the primordial spectrum (at $a> a_{\rm init}$) from its role in resolving the horizon
anomaly (at $a<a_{\rm trans}$). 

Thus, it appears that such proposed modifications to the conventional chaotic inflationary
scenario may satisfy at least three of the observational requirements. Unfortunately,
it does not seem likely that all four can be upheld simultaneously. As one can see
in Figure~\ref{fig:Fig2}, any modification to $V(\phi)$ consistent with the first
three conditions will fail to also satisfy the fourth. Though the value of $a_{\rm init}$
is at least partially selected to comply with the observed cutoff $k_{\rm min}$, 
there would clearly be modes with larger wavelengths ($\lambda_k>\lambda_{\rm max}$)
exiting the horizon prior to $a_{\rm init}$, e.g., at point 3 in the first de Sitter 
expansion. And these will have even larger amplitudes than they would have had if the 
slow-roll expansion had started prior to point 3. In other words, it
is challenging to justify an earlier phase of de Sitter expansion, beginning within 
the Planck regime, given that it would produce a fluctuation spectrum inconsistent
with the CMB data.

But were we to follow such a modification, we have for the KD phase (Region II)
\citep{LiuMelia:2020}:
\begin{equation}
H = M a^{-3}\;,\label{eq:KDavsH}
\end{equation}
\vfill\eject\noindent
where $M$ is a constant, and so
\begin{equation}
{{H_{\rm init}} \over {H_{\rm Pl}}}= \left({{a_{\rm trans}} \over {a_{\rm init}}} \right) ^3\;,\label{eq:KDavsH2}
\end{equation}
using the fact that $H$ remains roughly constant at its Planck value in Region I. But the first 
Friedmann equation yields 
\begin{equation}
{{H_{\rm init}} \over {H_{\rm Pl}}}= \left[{ {\rho(t_{\rm init})} \over {\rho(t_{\rm Pl})}} 
\right] ^ {1 \over 2}\;,\label{eq:KDrhovsH}
\end{equation}
and therefore using $a_{\rm init} \approx 9.7 \times 10^{-56}$ gives $a_{\rm trans} 
\approx 8.1 \times 10^{-58}$. The Hubble constant must drop by about 12 orders of magnitude
during this brief phase lasting an equivalent $\sim 5$ e-folds.

By assuming the expansion is de Sitter right from the Planck scale, this scenario mimics
the earliest moments of classic chaotic inflation, setting the initial conditions in the same
fashion. For example, since the scale factor $a(t)$ can be arbitrarily small during this
first de Sitter phase, the horizon problem can always be eliminated. All the modes will have
been seeded prior to the Planck time (as in the traditional inflationary concept), so the 
Bunch-Davies vacuum condition can (in principle) be satisfied, allowing the fluctuations 
to be properly normalized. Nevertheless, the long-standing trans-Planckian anomaly is still
present \citep{Martin:2001}. 

It would seem that unconventional modifications such as these may address some of
the challenges discussed in this paper. But several long-standing problems would still 
remain, including inflation's violation of the SEC and the trans-Planckian anomaly (see also
\citealt{Melia:2020b}, and references cited therein). And even so, it appears unlikely
that all the minimal observational requirements can be satisfied simultaneously.

A much more drastic modification to the basic inflationary paradigm would
be the introduction of alternative models, notably loop quantum cosmology 
\citep{Bojowald:2008,Ashtekar:2003,Ashtekar:2011,Martin-Benito:2009,Taveras:2008,Bhardwaj:2019,Navascues:2021}.
In these models, the big bang singularity is replaced with a bounce, where all 
the physical quantitites remain finite. In particular, the Hubble constant is
actually zero at the very beginning, implying an arbitrarily
large gravitational horizon---at least effectively---so that subsequent 
inflation does not seem necessary to solve a horizon problem in this 
restricted sense. The principle purpose of inflation is to produce
the ``correct" fluctuation spectrum, via the conventional mechanism of horizon
crossing and freezing, which would require slow-roll conditions over most of 
the (presumably) de Sitter expansion while the inflaton field is dominant. 

As we have noted earlier, the expansion in these models proceeds via a pre-inflationary
regime of kinetic dominance that eventually morphs into the more traditional 
slow-roll inflation. And the main effect of the loop quantum bounce on the
primordial perturbations translates into a large infrared suppression in their
spectrum, covering an even broader range of scales than one would see with a
kinetically-dominated pre-inflationary phase in the conventional picture.

The form and properties of this power spectrum, however, depend critically
on the specific choice of the vacuum state. In traditional inflation, one
merely assumes that the fluctuations were seeded far enough into the distant
past that their wavelengths were much smaller than the spacetime curvature
radius at that time. Born in what is effectively Minkowski space, these
fluctuations could therefore be quantized canonically with a time-independent
frequency. That is, one could invoke the existence of a Bunch-Davies vacuum
for the conventional picture, though our analysis in this paper now suggests
that even this approach is flawed for the new chaotic inflation. 

But when the mode frequencies are time dependent, as they are in loop
quantum cosmology, the Bunch-Davies vacuum is not an option. There is an
ambiguity about the time at which the modes should be canonically quantized.
Strides made in recent years \citep{Ashtekar:2020,Navascues:2021} have pointed
to a possible resolution of this problem, with an indication that the fluctuations
should be seeded at the time of the bounce itself.

Such alternative cosmologies may resolve some of the challenges imposed on the 
inflationary paradigm by the current observations. We point out, however, that 
much work would still need to be done to bring these models into full compliance 
with the data. A glaring example is provided by the state-of-the-art 
angular correlation function predicted by the version of loop quantum cosmology
proposed by \cite{Ashtekar:2020}, which clearly improves upon the poor fit of the 
CMB data provided by the standard picture, but yet clearly also does very poorly 
compared to a simple, hard cutoff, $k_{\rm min}$, in the primordial spectrum. 
One can easily see this by inspection and a comparison of the $C(\theta)$ curves 
in Figure~3 of \cite{Ashtekar:2020} and Figure~2 of \cite{MeliaLopez:2018}.

\section{Conclusion}\label{conclusion}
The emergence of a cutoff $k_{\rm min}$ in the primordial power spectrum, implying
a delayed initiation time for inflation,  affirms the view that the conventional 
quasi de Sitter expansion could not have started until well after the Planck time
in order to avoid a conflict with the required initial energy density in the
cosmic fluid. In this paper, we have demonstrated that this new picture faces
several serious challenges. The first arises from the conflict in solving the
horizon problem while simultaneously fitting the measured $k_{\rm min}$ value.
Second, the initial conditions of inflation are now unspecified, which may be
solved during a pre-inflationary phase, but then raising the bigger question
of why we would even need inflation to solve the horizon problem in the first
place. Finally, this new inflationary scenario fails to provide a mechanism
for properly quantizing the seed fluctuations.

Of course, an important caveat in this discussion is that we have focused largely
on slow-roll models, with only a limited range of possible pre-inflationary
expansions. Perhaps a more inventive approach can circumvent at least some
of these difficulties, as we have illustrated with the simple toy model in
\S~\ref{discussion}. In the end, however, the preservation of
inflation may require an entirely new cosmological framework. Or inflation 
may turn out to be unnecessary after all \citep{Melia:2013c,Melia:2020}, and 
we shall learn that the challenges it now faces are an indication that it simply 
never happened.

\begin{acknowledgments}
We are grateful to the anonymous referees for their helpful
comments that have led to several significant improvements in the presentation 
of this material. 
\end{acknowledgments}

\vfill\newpage
\bibliographystyle{aasjournal}
\bibliography{ms}{}

\end{document}